\documentclass[12pt,preprint]{aastex}

\usepackage{emulateapj5}
\usepackage{apjfonts}

\shorttitle{Constraining jet theories}
\shortauthors{Alon Retter}

\begin{document}


\title {Constraining jet theories using nova outbursts}

\author{Alon Retter\altaffilmark{1}}


\altaffiltext{1}{Department of Astronomy and Astrophysics, Pennsylvania 
State University, 525 Davey Lab, University Park, PA 16802-6305; 
retter@astro.psu.edu}







\begin{abstract}

It is commonly accepted that jets have not been observed in cataclysmic 
variables (CVs) so far. This absence was recently explained by their low 
mass transfer rates compared with objects with jets. A mass accretion 
limit for jets in CVs was proposed to be $\dot{M}\sim10^{-7}-10^{-6}$ 
M$_{\odot}$/yr. There was, however, a report of evidence for jets in 
V1494~Aql, the second nova that erupted in Aquila in 1999. In this work, 
we estimate the mass transfer rate of this system around the reported 
event and show that it is consistent with the above theoretical limit 
for jets. We further propose that the X-ray flare that was observed in 
this object may be connected with a jet as well. The appearance of jets 
in novae is actually expected since during the early decline from outburst 
some are alike supersoft X-ray sources, in which jets have been found. 
The detection of jets in novae also fits the suggestion that in 
addition to the presence of an accretion disk, a hot central source is 
required for the formation of jets. The observations of jets during the
transition phase in V1494~Aql can be regarded as evidence for the early
existence of the accretion disk in the system. This conclusion supports 
our previous suggestion for a link between the transition phase in novae 
and the re-establishment of the accretion disk. We further speculate that 
jets may be restricted to transient novae. If our ideas are confirmed, 
jets should be common in transient novae and may be formed several times 
during the transition phase and perhaps even long after it ends. 
In classical novae jets may be launched and observed in real time. As 
binary systems, these objects are easy to study. Novae may, therefore, 
be key systems in understanding the formation and evolution of jets 
and ideal targets to test and constrain jet theories. Adding novae to 
the classes of systems showing jets should be very exciting and 
rewarding.




\end{abstract}


\section{Introduction}


Jets are one of the most spectacular astrophysical events. A jet is 
a narrow beam of matter or radiation emerging from an astronomical
object. Usually two beams at opposite directions are observed from a 
central source. Collimated jets have been detected in young stellar 
objects, low mass and high mass X-ray binaries, supersoft X-ray sources, 
planetary nebula nuclei, symbiotic stars and active galactic nuclei. 




Livio (2000) reviewed the observations and theory of jets and argued 
that the data of systems showing jets are consistent with the presence 
of accretion disks in all cases. Using a phenomenological approach, 
he also suggested that jets are launched from the center of the disk. 
It is commonly assumed that magnetic fields play a crucial role in the 
formation of jets and that they are accelerated and collimated 
hydromagnetically. Livio (2000) further proposed that the production 
of jets may require an additional heat or wind source probably 
associated with the central object. Yet, the jet mechanism is still 
obscure and so far numerous theories have been proposed to explain 
their formation.



 
Cataclysmic variables (CVs) contain a white dwarf star and a low mass 
red dwarf companion which fills its Roche lobe.  Typically, the white 
dwarf accretes mass from an accretion disk that surrounds it.  In most 
cases, jets have been associated with accretion disks, but it is widely 
accepted that jets have not been observed in CVs so far (e.g., Sokoloski 
et al. 2004; Hillwig, Livio \& Honeycutt 2004).  This latter perception 
is somewhat surprising, especially since jets have been detected in many 
symbiotic systems which contain accreting white dwarfs (Sokoloski et al. 
2004; Galloway \& Sokoloski 2004). It should be noted that an earlier 
report of jets in a CV (T Pyx) was based on a misinterpretation of the 
data (O'Brien \& Cohen 1998), so to date there has not been any 
convincing evidence of jets in CVs.



CVs can be divided into two main subgroups: dwarf novae and nova-like 
systems. Dwarf novae have relatively low mass transfer rates
($\dot{M}\sim 10^{-11}-10^{-10}$ M$_{\odot}$/yr) and their accretion 
disks are thermally unstable leading to dwarf nova outbursts typically 
every several months in their light curves. Nova-like systems have 
higher accretion rates ($\dot{M}\sim 10^{-9}-10^{-8}$ M$_{\odot}$/yr) 
resulting in stable accretion disks and no dwarf nova eruptions. After 
the white dwarf accretes enough material on its surface, and once its 
temperature and density reach some critical values, a 
thermo-nuclear-runaway occurs, giving rise to the observed nova 
outburst. It is believed that all CVs have nova eruptions every 
several ten thousands years (Warner 1995). 

Soker \& Regev (2003) pointed out recent observations that showed no 
difference in the X-ray properties of young stellar objects with or 
without jets (Getman et al. 2002). They, thus, argued that the observations
are inconsistent with models that invoke the presence of magnetic fields 
for the launching mechanism of jets. Instead, they advanced a thermal 
model for the formation of jets in young stellar objects. It states that 
the accreted matter is shocked because of large gradients of physical 
quantities in the boundary layer near the accreting star. Soker \& Lasota 
(2004) further applied this model to CVs to explain the absence of jets 
in these systems. They derived a rough estimate to the limiting mass 
transfer rate for jets in CVs: $\sim 10^{-7}-10^{-6}$ M$_{\odot}$/yr. 
As CVs typically accrete at much lower rates ($\dot{M}\la
10^{-8}$ M$_{\odot}$/yr -- e.g., Knigge \& Livio 1998), they should not 
show jets. 



\subsection{The transition phase in classical novae}

After the nova outburst, the white dwarf is quite hot and its cooling 
down process takes about a decade (Prialnik 1986). The nova should have a 
high accretion rate that would gradually decrease in time. Classical novae 
may thus be excellent objects to check the evolution of jets if indeed 
they are formed in these systems. The accretion disk may be destroyed 
during the nova blast resulting in no accretion at all. However, strong 
evidence for the presence of the disk in novae several months-years 
after eruption have been found (Retter, Leibowitz \& Ofek 1997; 
Skillman et al. 1997; see also Leibowitz et al. 1992; Retter, Leibowitz 
\& Kovo-Kariti 1998). In addition, the white dwarf heated by the nova 
outburst may supply the extra energy source required for the formation 
of jets (Livio 2000). These systems may therefore, be ideal laboratories 
to test jet theories. The contribution from the white dwarf can be 
removed using theoretical models of its cooling (Prialnik 1986; Somers 
\& Naylor 1999). The observations or non-detection of jets at different 
stages of the nova decline can put strong constraints on jet models. 



Classical novae can be broadly catalogued into three major subclasses 
according to the shape of their light curves. They either show a 
smooth decline, a deep minimum or oscillations during the transition 
phase several weeks-months after maximum light. The deep minimum in the 
light curve is explained by the formation of dust that may be restricted 
to systems with CO white dwarfs compared with ONeMg primaries in other 
novae (e.g., Rudy et al. 2003). The nature of the oscillations during 
the transition phase is, however, still under intensive debate.


Retter (2002) reviewed this issue and proposed a possible connection 
between the transition phase in classical novae and the re-formation 
of the accretion disk in a subclass of CVs called intermediate polars. 
These are systems whose inner part of the accretion disk is truncated 
by a moderate magnetic field of the white dwarf instead of a strong 
magnetic field, as in the case of polars. According to this idea, in a 
nova outburst it is easier to destroy the disk in intermediate polars 
as the inner part of the disk is missing. Retter (2002) thus proposed 
that the re-establishment of the accretion disk and the interaction 
between the inner part of the disk with the magnetic field of the white 
dwarf may form the oscillations observed during the transition phase. 
The test of this idea is very simple: novae with a transition phase 
(hereafter transient novae) should be intermediate polars and have fast 
spin periods. The detection of short-term periodicities in the X-ray 
light curves of two recent transient novae, V1494~Aql 1999 (Drake et 
al. 2003) and V4743~Sgr 2002 (Ness et al. 2003) is consistent with 
this suggestion. We note, however, that Drake et al. (2003) interpreted 
the variations in V1494~Aql as stellar pulsations of the hot white dwarf, 
while Ness et al. (2003) discussed both models as the cause for the X-ray 
fluctuations in V4743~Sgr.

In this work we investigate the first reliable report of jets in CVs, in
V1494~Aql, which is a transient nova.



\subsection{Nova V1494~Aql 1999}

The outburst of V1494~Aql was discovered in 1999 December 1 (Pereira 
1999). Reaching V$\sim$4 at maximum a day later, it was one of the 
brightest novae in the century. Figure 1 presents the visual light curve
of V1494~Aql during the first couple of years following discovery. A 
few months after the outburst the light curve was characterized by 
$\sim$1-2 mag oscillations with a quasi-period of 1-2 weeks 
typical of the transition phase in novae (Kiss \& Thomson 2000; Iijima 
\& Eseno\u{g}lu 2003). 



\begin{figure*}
\epsscale{1.7} 
\plotone{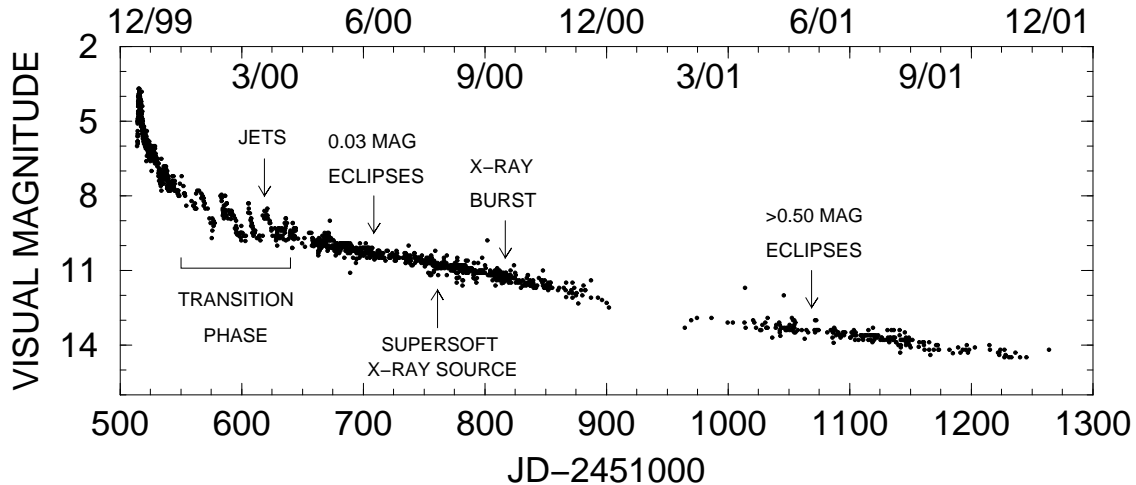}
\caption{The visual light curve of V1494~Aql during the first two years 
after discovery, taken from the VSNET (Variable Star Network). The nova 
had 1-2 mag oscillations with a quasi-period of 1-2 weeks during the 
transition phase, several months after peak outburst. Evidence for jets 
was found in 2000 March 16-17. V1494~Aql was classified as a supersoft 
X-ray source in 2000 August 6. An X-ray burst was observed in the system 
in 2000 October 1. Eclipses with a period of 3.2 h were first detected 
in the optical light curve of the nova in 2000 June with an amplitude of 
0.03 mag. In 2001 June, the eclipse depth was much larger.}
\end{figure*}




Extensive photometric observations of V1494~Aql revealed a periodic 
signal at 3.2 h (Retter et al. 2000; Bos et al. 2001; Pavlenko, Dudka 
\& Baklanov 2003; Barsukova \& Goranskii 2003; Kato et al. 2004). 
The folded light curve showed eclipses whose physical nature is 
still unknown. Bos et al. (2001) proposed that the accretion disk is 
eclipsed by the secondary star. Pavlenko et al. (2003) argued that 
the periodic variations can be explained by the presence of a strong 
magnetic field of the white dwarf and no accretion disk (i.e., polar 
model). Barsukova \& Goranskii (2003) favored an intermediate polar 
model for the nova, with a truncated disk and a moderate magnetic 
field of the white dwarf. Recently, Hachisu, Kato \& Kato (2004) 
demonstrated that the orbital light curve can be understood by the 
formation of a spiral structure in the accretion disk. 

Iijima \& Eseno\u{g}lu (2003) analyzed spectroscopic observations of 
V1494~Aql during its early fading phase. Their data were intermittently 
taken during different stages of the decline from outburst, including 
the transition phase. Around one of the extrema of the long-term 
oscillations, two spectra showed symmetric emission features around the 
H$\alpha$ and H$\beta$ lines with a velocity of about 2800 km/s. This 
was interpreted by the authors as evidence for jets. In this work, we 
assess this suggestion and discuss its implications. 


\section{The mass transfer rate in V1494~Aql}

The presence of jets in V1494~Aql can be supported by estimating its 
mass accretion rate during the `jet event' and showing that it is 
consistent with the theoretical limit. First, we assume that an accretion 
disk existed in this system in 2000 March. This assumption is justified 
by the fact that in most CVs the white dwarf accretes material from its 
companion star through a disk and strong evidence for its existence in 
novae several months-years after outburst has been detected (Section~1.1). 
In 2000 March, at the minimum of the oscillation, V1494~Aql was at 
V$\sim$10 (Fig.~1), about 6 mags fainter than the main maximum of the 
outburst. Evidence for the presence of the accretion disk was first 
found in Nova V1974 Cyg~1992 when it was about 8 mags below maximum 
(Retter \& Leibowitz 1998). The disk may have been present earlier in 
this system, although not as the dominant light source. Thus, it seems 
reasonable to assume that the accretion disk existed in V1494~Aql during 
the `jet event'. Conversely, we can claim that the detection of jets in 
V1494~Aql is an indication for the presence of the accretion disk in 
this nova in 2000 March.

The eclipses in V1494~Aql were discovered in 2000 June with an amplitude
of 0.03 mag (Fig.~1; Retter et al. 2000). A year later, the eclipse 
depth was about 0.5 mag (Bos et al. 2001), and actually should be even 
larger after subtracting the contribution of light from the nearby 
background star (Kiss, Cs\'ak \& Derekas 2004). Thus, in 2001 June the
accretion disk was probably the main optical source.

The mass transfer rate of V1494~Aql will be estimated from its physical 
parameters. To simplify this calculation, we assume that the visual 
luminosity of V1494~Aql was completely dominated by the accretion disk 
in 2001 June and that the disk was not irradiated by the hot white dwarf. 
We follow Webbink et al. (1987) who developed a formula for the mass 
accretion rate of a CV as a function of its visual magnitude, distance, 
interstellar reddening and inclination angle. The visual magnitude of 
the nova in 2001 June was V$\sim$13.5 (Fig.~1; Kiyota, Kato, \& Yamaoka 
2004). The distance to V1494~Aql and the interstellar extinction are 
taken as d=1.6 kpc and A$_{V}$=2.0 (Iijima \& Eseno\u{g}lu 2003). The 
presence of eclipses constrains the inclination angle to high values, 
and it is assumed to be i=80$^\circ$. These values yield a mass transfer 
rate of about 5$\times 10^{-6}$ M$_{\odot}$/yr. 


The correct mass accretion rate of V1494~Aql in 2001 June should be 
somewhat lower than this value because of the assumption that the 
contribution from the white dwarf is negligible. The calculation shows, 
however, that if the accretion disk already existed at this time, the 
mass transfer rate was very high. If we further assume that during the 
`jet event' in 2000 March, the mass accretion rate was of the same order 
(and it may have even been higher because the nova was brighter), then 
it obeys the critical limit for jets in CVs set by Soker \& Lasota (2004).


Another independent way of estimating the accretion rate is to postulate 
that in the transition phase the accretion disk is destroyed during 
each oscillation, which we interpret as a mini-outburst (see below). 
An accretion disk in a typical dwarf nova before outburst has a mass 
of $\sim 10^{-9}$ M$_{\odot}$ (e.g., Cannizzo, Gehrels \& Mattei 2002). 
The mass of the disk in old novae and nova-like systems (as should be 
in V1494~Aql when it returns to quiescence) is probably even larger. 
Assuming that the disk is re-established between two successive 
mini-outbursts, thus within a few days, this is translated to mass 
transfer rates of the order of 10$^{-7}$ M$_{\odot}$/yr. This value 
should be regarded as a low limit to the mass accretion rate of the 
nova in 2000 March.

As a summary of this section, we conclude that if an accretion disk
existed in V1494~Aql in 2000 March, then its mass transfer rate was
probably in the range 10$^{-7}$--5$\times 10^{-6}$ M$_{\odot}$/yr,
say $\sim 10^{-6}$ M$_{\odot}$/yr.


\section{Discussion}

We feel secure to argue that V1494~Aql had an accretion disk during the 
jet event and that the mass transfer rate at that time was relatively 
very high and almost certainly above the theoretical limit for jets 
despite the uncertainties in the calculations made in the previous section. 
To support this claim, we note that some novae have a supersoft X-ray 
stage for several months-years during the early decline from outburst 
(e.g., Orio, Covington \& \"{O}gelman 2001; Ness et al. 2003). Supersoft 
X-ray sources have high accretion rates and jets have been observed 
in a few systems (see references in Livio 2000; Soker \& Lasota 2004). 
V1494~Aql was classified as a supersoft X-ray source in 2000 August 6 
(Drake et al. 2003). Therefore, the presence of high mass transfer 
rates and jets in this nova during its early fading phase is consistent 
with the observations of these systems as well.

On 2000 October 1, the X-ray emission from V1494~Aql increased suddenly 
by a factor of $\sim$10 for about 15 minutes (Drake et al. 2003). This 
phenomenon is still unexplained. Our first estimate of the mass 
transfer rate in V1494~Aql ($\dot{M}$$\sim$5$\times10^{-6}$ 
M$_{\odot}$/yr) is for 2001 June, eight months after this event. In 
2000 October, the mass accretion rate should have, therefore, been 
pretty large. We thus propose that this flare or burst might be 
connected with the formation of a jet as a result of a high mass 
transfer event. 


Recent optical observations of Nova V4745~Sgr 2003 showed that spectra 
of this transient nova during the major maximum of its eruption and 
other maxima of the oscillations during the transition phase were 
alike. This discovery invoked an explanation of several repeating 
thermo-nuclear mini-outbursts as the cause of the oscillations 
(Cs\'ak et al. 2004). We note that these observations seem inconsistent 
with the explanation of the transition phase by winds (Shaviv 2001).
Therefore, we refine Retter's (2002) model (Section~1.1) and consider 
the following scenario for the transition phase: the nova outburst 
destroys the accretion disk in intermediate polars. After the white 
dwarf cools down and once its radiation pressure falls below a certain 
value, the disk is re-established within a few days, which leads to 
instantaneous high mass transfer rate. This triggers another mini nova 
outburst, that should demolish the disk again. The cycle may repeat 
several times with decreasing amplitudes until the mass transfer 
rate and the white dwarf temperature are too low to trigger another 
mini-outburst. 



We suggest then that jets should be present during the transition phase 
in classical novae when presumably the accretion disk is re-established 
and the mass accretion rate is high enough, above the theoretical 
limit. They should be formed several times in accord with the number 
of mini-outbursts. In addition, if the mass transfer rate is still 
large enough, above the critical value, collimated jets may be present 
even after the transition phase ends. The formation of jets should be 
ceased once the mass transfer rate drops below the critical limit. 
When the nova returns to quiescence, it should behave like a typical 
CV and not show jets. 


The formation of jets has been linked with the presence of magnetic 
fields of the accretion disk itself or of the accreting star. Following 
our suggestion that the transition phase occurs in intermediate polars 
and assuming that jets are restricted to transient novae, we propose 
that the requirements for jets are the presence of an accretion disk 
and a moderate magnetic field of the inner source rather than the disk. 
This suggestion will be tested by future observations.
 

The thermal launching model presented by Soker \& Lasota (2004) requires 
that the accretion disk be strongly disturbed at its inner part. This 
can be achieved by a boundary layer with a star, or by a strong stellar 
magnetic field (see second sentence in section 1.3 of Soker \& Regev 
2003). Therefore, the discussion in the present paper is valid for 
jets launched by systems where the disk inner radius is at the stellar 
surface as well as for jets from intermediate polars, where the disks are 
truncated by a moderate white dwarf magnetic field. 


\section{Summary and Conclusions}





We find that during its jet event V1494~Aql had a high mass accretion 
rate, $\sim$10$^{-6}$ M$_{\odot}$/yr, consistent with the theoretical 
limit for jets in CVs (Soker \& Lasota 2004). It is argued further 
that the peculiarities of V1494~Aql and V4743~Sgr in the X-ray may 
be explained by the ideas presented in this paper. According to our 
scenario when the nova decays and the accretion disk is reformed, jets 
should appear as a consequence of the temporary high accretion rate. 
The presence of jets in novae shortly after eruption seems natural 
because novae typically have a supersoft X-ray phase during the decline 
from outburst and jets have been detected in supersoft X-ray sources.


The evidence for jets in classical novae requires further support and it 
may be a coincidence that they appeared in a transient nova, nevertheless 
we speculate that jets may be restricted to transient novae. Furthermore, 
we propose that this phenomenon may be connected with the re-formation of 
the accretion disk and thus may be repeated several times before the mass 
transfer rate drops below the critical limit for jets. 

We suggest that novae are ideal targets for the study of jets. This 
is because there is an evolution from high to low accretion rates 
within a few months or years in these systems in addition to a 
gradual change of the dominant light source from the hot white dwarf 
to the accretion disk. Further observations or non-detection of jets 
at different stages of the nova decline can put strong constraints on 
jet models. This paper thus calls for extensive X-ray, radio and optical
spectroscopic monitoring of classical novae (and especially of transient 
novae) during their early decline stage. In addition, we encourage 
researchers to look for evidence of jets in archive high resolution 
spectroscopic data of novae.



The addition of classical novae to the known groups of objects showing 
jets should be quite exciting and may yield new insights on the formation
of jets. It is very likely that novae, which are easy to study as binary 
systems, will supply the missing information on the jet mechanism.

\acknowledgments

 

We warmly thank M.T. Richards for many useful comments on an early 
draft of the paper and N. Soker for his encouragement to publish this 
work and for helpful discussions. This work was partially supported by 
a postdoctoral fellowship from Penn State University.




\bibliography{%
/home-astron/retter/bib/mnemonic,%
/home-astron/retter/bib/mnemonic-simple,%
/home-astron/retter/bib/all_algol%
}

\bibliographystyle{/home-astron/bedding/bstinputs/natbib/mynatbib}

\end{document}